\documentclass[10pt,conference]{IEEEtran}

\usepackage{graphicx}   
\usepackage{url}        
\usepackage{amssymb}
\usepackage{amsmath}    
\usepackage{hyperref}
\usepackage[export]{adjustbox}
\usepackage{subcaption}
\usepackage{fancyhdr}

\fancypagestyle{specialfooter}{%
\fancyhf{}

\fancyfoot[R]{ \noindent\fbox{%
\parbox{\textwidth}{%
{\footnotesize This work has been submitted to the IEEE for possible publication. Copyright may be transferred without notice, after which this version may no longer be accessible.}
}
}}
}
\begin{document}

\title{Pooling techniques in hybrid quantum-classical convolutional neural networks}
\author{
\IEEEauthorblockN{Maureen Monnet\IEEEauthorrefmark{1}, Hanady Gebran, Andrea Matic-Flierl\IEEEauthorrefmark{1}, Florian Kiwit,}
\IEEEauthorblockN{Balthasar Schachtner\IEEEauthorrefmark{2}, Amine Bentellis\IEEEauthorrefmark{1}, Jeanette Miriam Lorenz\IEEEauthorrefmark{1}}
\IEEEauthorblockA{\IEEEauthorrefmark{1}Fraunhofer Institute for Cognitive Systems IKS,  Munich, Germany}
\IEEEauthorblockA{\IEEEauthorrefmark{2}LMU University Hopsital, Munich, Germany}
\{maureen.monnet, andrea.matic-flierl, amine.bentellis, jeanette.miriam.lorenz\}@iks.fraunhofer.de \\ balthasar.schachtner@med.lmu.de
}
\maketitle\thispagestyle{specialfooter} 

\begin{abstract}
Quantum machine learning has received significant interest in recent years, with theoretical studies showing that quantum variants of classical machine learning algorithms can provide good generalization from small training data sizes. However, there are notably no strong theoretical insights about what makes a quantum circuit design better than another, and comparative studies between quantum equivalents have not been done for every type of classical layers or techniques crucial for classical machine learning. Particularly, the pooling layer within convolutional neural networks is a fundamental operation left to explore. Pooling mechanisms significantly improve the performance of classical machine learning algorithms by playing a key role in reducing input dimensionality and extracting clean features from the input data. In this work, an in-depth study of pooling techniques in hybrid quantum-classical convolutional neural networks (QCCNNs) for classifying 2D medical images is performed. The performance of four different quantum and hybrid pooling techniques is studied: mid-circuit measurements, ancilla qubits with controlled gates, modular quantum pooling blocks and qubit selection with classical postprocessing. We find similar or better performance in comparison to an equivalent classical model and QCCNN without pooling and conclude that it is promising to study architectural choices in QCCNNs in more depth for future applications.
\end{abstract}


\section{Introduction}
\label{sec:intro}

The recent years have seen an increased interest in quantum machine learning (QML) (\cite{Wittek2016}; \cite{Schuld2021}; \cite{Ganguly2021}). QML is a discipline at the intersection of machine learning and quantum computing (QC), where in context of this paper we are interested in quantum-enhanced machine learning algorithms processing classical data like images. However, the practical benefits of these algorithms remain unclear. A common approach to implement QML algorithms is via variational quantum circuits (VQCs)~\cite{Cerezo2021}, which consist of a circuit architecture with parametrized gates.


Current Noisy Intermediate-Scale quantum (NISQ) computers feature a limited number of $\mathcal{O}(100)$ qubits, with limited connectivity and limited gate fidelities. Therefore, hybrid QML algorithms able to deal with shallow quantum circuit sizes and only requiring a small number of qubits are most promising for near-term applications of QC. In these hybrid algorithms, an iterative interaction between classical computers and QC occurs. As QC will likely be combined with classical computers as additional quantum processing unit (QPU), the QPU should ideally only be used for the parts of a calculation which are
significantly accelerated by the use of QC or are not possible
to calculate on classical computers. For example, a quantum computer may be used for the execution of variational parts of the QML algorithm, while a classical computer may calculate the parameter values of the variational part by minimizing a cost function.


Hybrid variants have been proposed for many well-established algorithms of classical machine learning, including convolutional neural networks (CNNs). CNNs are widely applied in computer vision tasks, such as image classification. They consist of convolutional layers, with which the characteristic features in the images are extracted, followed by fully connected layers used for classification. Commonly, a convolutional layer is followed by a pooling layer, where the extracted feature maps are down-scaled to a smaller data size. Thereby the feature maps become more robust to translations of the input and overfitting can be prevented. In principle, the whole CNN model can be translated to a quantum algorithm, but the size of the resulting quantum algorithm is likely to require quantum hardware beyond the presently available NISQ devices. If instead just certain parts of a CNN are moved to QPU, the resulting algorithm can already be NISQ-compatible. Particularly, mapping the convolutional or the pooling layers to a quantum variant represents a promising approach in terms of the number of available qubits in NISQ-devices, as these layers process the data sequentially in smaller slices, in contrast to the linear layers, which operate on all features simultaneously. While the usage of quantum convolutional layers was extensively studied in recent research, the effect of quantum pooling layers is yet relatively unexplored.

In this work, we test the suitability of quantum pooling operations for a medical imaging task. In particular, we analyze 2D ultrasound images of the breast to identify malign lesions. For this classification task, we investigate four different techniques for performing a quantum and hybrid quantum-classical pooling operation, namely the use of 1) mid-circuit measurements, 2) auxiliary qubits with controlled gate operations, 3) modular quantum pooling blocks, and lastly 4) qubit selection with a non-linear classical postprocessing function. We compare their performance with the one of a quantum convolutional operation as well as a fully classical approach. 

Our contributions are thus threefolds:
\begin{itemize}
\item We design and compare 4 different quantum pooling architectures for hybrid quantum-classical convolutional neural networks with respect to the training and validation performance on ultrasound images to identify malign lesions of the breast.
\item We further contrast these architectures with both a classical convolutional neural network and a quantum-classical convolutional neural network without pooling containing a similar number of trainable parameters.
\item We seek to provide an explanation for the performance of our developed models using a quantum relevant metric.
\end{itemize}
The paper is structured as follows. We present the related work in section~\ref{relatedwork}. We introduce the chosen dataset as well as the background information relevant to all designed hybrid quantum-classical convolutional neural networks and to the classical convolutional neural network in section~\ref{background}. We define all tested quantum pooling architectures in section~\ref{section4} and present and interpret their performance. We introduce a quantum metric in section~\ref{section5}, the effective dimension, and investigate its correlation with training performance, before coming to conclusion in section~\ref{ccl}.

\section{Related Work}\label{relatedwork}

The works in (\cite{cong2019quantum},\cite{kerenidis2020quantum},\cite{lu2021quantum},\cite{wei2021quantum},\cite{li2021quantum},\cite{hur2022quantum}) studied the possibility of moving all components of a CNN on a QC. Since this is generally not feasible with current NISQ-devices, hybrid quantum-classical CNNs (QCCNNs) have been introduced, in which only parts of a CNN are executed on a quantum computer. In (\cite{henderson2020quanvolutional}, \cite{mattern2021variational}, \cite{liu2021hybrid}) the first classical convolutional layer is replaced by either an untrainable or trainable quantum convolutional layer. Using this architecture, (\cite{henderson2020quanvolutional},\cite{mattern2021variational}) achieved a good performance on the MNIST dataset \cite{lecun1998gradient}. We have extended that work in \cite{matic2022quantum}, where we have proposed different architectures, also using different encoding schemes and variants of VQCs, on multiple medical datasets. In particular, we also compared their performance to fully classical CNNs and thereby demonstrated a good performance of hybrid QCCNNs. 

QCCNN architectures may profit from the inclusion of pooling layers, since those lead to a dimension reduction of the feature maps, and consequently reduce the number of parameters to learn, potentially preventing overfitting. Different proposals have been made in this direction. In (\cite{cong2019quantum, lu2021quantum, zheng2021}) the quantum pooling operation is achieved through mid-circuit measurements, which means that measurements are performed for only a subset of the qubits in the quantum circuit. Depending on their results, different unitary operations are applied on the adjacent qubits. In general, mid-circuit measurements represent a promising method for quantum applications, where a reduction of the data is necessary. For example, the work in \cite{botelho2022error} explores their usage for error mitigation. 
In \cite{wei2021quantum} quantum pooling layers are realized by abandoning a subset of qubits in the quantum circuit, i.e. their quantum states are not measured and further processed in the algorithm. Similarly, in \cite{Hur:2021zyz} quantum pooling is achieved through parametrized two-qubit operations, after which the state of one of the two qubits is ignored. 

The aforementioned work on pooling layers were all proposed for full quantum variants of CNNs. Inspired by some of these ideas, we investigate further techniques to realize and compare quantum poolings in hybrid QCCNNs. For this, we build upon our preceding studies in \cite{matic2022quantum}, where we studied the effect of replacing a classical convolutional layer by a quantum variant in medical image classification tasks.

\section{Background}\label{background}

In this section, we will explain the dataset used in our studies and the necessary background information related to our experiments. This includes the design of both the classical CNN and the hybrid variants. For the latter, we focus here only on the quantum convolutional layer as well as on aspects common to all tested QCCNNs in this work. The pooling architectures are discussed in detail in section~\ref{section4}.

\subsection{Dataset}

Work by \cite{Caro2021} has shown that certain QML algorithms like the quantum-variant of CNNs are capable to generalize well even when little training data is available. In medical imaging, the acquisition of large high-quality datasets is often expensive and difficult because it requires careful expert annotation. Additional challenges come from high imaging costs and privacy concerns. Therefore, medical datasets are usually relatively small \cite{Prevedello012019}, common dataset sizes are of the order of 100 to 1000 images. To ensure an accurate and reliable performance of artificial intelligence (AI) algorithms in medical imaging, one thus needs to find an efficient way to process such small and complex imaging data \cite{Willemink2020}. Therefore, medical imaging is an interesting area to be explored with QML.

In this work, the performance of all developed QCCNN models is explored on the BreastMNIST dataset from the MedMNIST collection \cite{yang2021medmnist}\cite{al2020dataset}. This dataset is characterized by a considerably low number of training samples as is typically the case in medical imaging tasks and consists of 546 training and 78 validation images of breast ultrasounds images, which are preprocessed as described in \cite{al2020dataset} and downsampled to a $28 \times 28$ resolution. The task is a binary classification between non-malignant - a combination of benign and normal states - and malignant lesions.

\subsection{Classical CNN and QCCNNs in this work}
\label{section3}

As classical baseline, we use the architecture presented in our previous work in~\cite{matic2022quantum}. This consists of a  convolutional layer with four filters of size $2 \times 2$, directly followed by a fully connected layer.

\begin{figure*}
    \centering
    \includegraphics[width=0.75\textwidth]{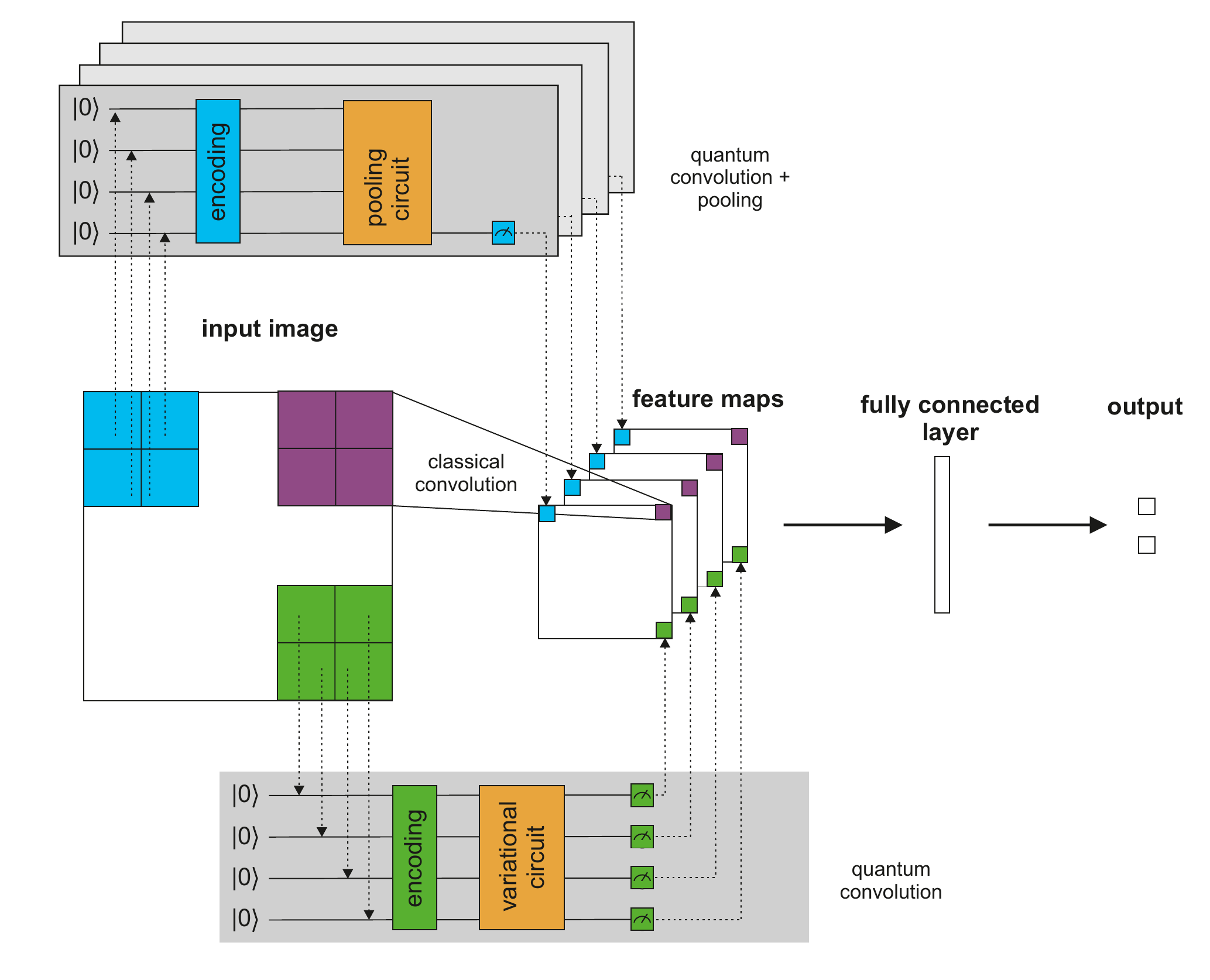}
    \caption{Architecture sketch of the CNN and QCCNNs with quantum convolutional layers with and without pooling.}
    \label{fig:img1}
\end{figure*}

In the QCCNNs, the convolutional layer is replaced by either a quantum convolution or by a combined quantum convolution + pooling operation to represent filters of size $2 \times 2$ moving over the image, as shown in figure~\ref{fig:img1}. The fully connected layer remains classical. Across all investigations conducted in the present study, a similar number of trainable parameters between the CNN and the QCCNNs is obtained with these architectures, making the comparison between fully classical and hybrid quantum-classical variants possible.

The quantum convolutional (+ pooling) layers consist of an encoding layer followed by a VQC and a measurement. The VQC is trainable and consists of a collection of trainable rotation gates and entangling two-qubit operations. For all QCCNNs designed in this work, the same encoding and measurement technique is employed: 

\begin{itemize}

\item Encoding: We use the so-called higher order encoding \cite{abbas2021power} first introduced in \cite{havlicek2019supervised}, as it is expected to work best thanks to its entangling capabilities. The qubits initially in state 0 are transformed using a Hadamard gate and a $Z$-axis rotation $R_{Z}\left(\pi * x_{n}\right)$, where $x_{n}$ denotes the $n$-th input. An entangling operation $R_{Z Z}\left(\phi_{i j}\right)$ is then applied to every possible qubit pair $i$ and $j$. This operation consists of a CNOT gate, a rotation $R_{Z}\left(\phi_{i j}\right)$, and another CNOT gate. The rotation $R_{Z}\left(\phi_{i j}\right)$ is applied on the $j$-th qubit and we use $\phi_{i j}=\pi * x_{i} * x_{j}$. The input data being normalized between $-1$ and $1$, the rotation gates in the encoding are operating between $-\pi$ and $\pi$.

\item Measurement: The expectation value of the Pauli-Z operator is computed for the desired qubits and the values are stored in individual feature maps, following the strategy of \cite{mattern2021variational}. With a convolutional filter size of $2 \times 2$ and the encoding mapping one input value to one qubit, we therefore use 4 qubits and hence produce 4 feature maps in this architecture.

\end{itemize}

For the VQC in the quantum convolutional layer, a basic entangling layer is chosen as baseline for comparison with the QCCNNs with quantum pooling, as it is the best performing variant for this use case in our previous work \cite{matic2022quantum}. In this instance, each qubit $i$ is rotated by a trainable angle $\theta_{i}$ around the $X$-axis before applying a sequence of CNOT entangling gates on adjacent qubit pairs.

Experiments are performed using PyTorch \cite{paszke2019pytorch} and the quantum circuits are simulated with Pennylane \cite{bergholm2018pennylane} while neglecting any noise effects. All networks, both classical and hybrid, are trained for 20 epochs with the Adam optimizer, a learning rate of 0.001 and a batch size of 8. A total of three runs, differing between each other in the initial parameters in the network, are executed per architecture to produce variance bands. No extensive hyperparameter tuning was done in this study due to the prohibitively long training times of QCCNNs.


\section{QCCNNs with a quantum pooling layer}\label{section4}

In the QCCNNs with a quantum pooling layer, the quantum convolution + pooling layer consists of data encoding with the higher order encoding described in section \ref{section3}, followed by one of the quantum pooling circuits detailed below. Only one qubit out of the four is then measured, as seen on figure \ref{fig:img1}. The quantum pooling layer is followed by a fully connected layer.

This quantum convolution + pooling operation yields one value as opposed to the quantum convolution where four values are returned. To achieve a fair comparison with the classical CNN and the QCCNN baseline described in section \ref{section3}, we therefore use four quantum kernels in parallel to form the quantum pooling architecture. This allows to maintain the same parameter count in the classical layers in all possible configurations. Moreover, this offers the advantage that four quantum kernels in parallel may allow to extract different features of the image. It is worth noting that the quantum pooling operation itself may be trainable. This is the case in pooling methods which comprise rotation gates, such as the pooling with mid-circuit measurements or the modular quantum pooling blocks.

\subsection{Mid-circuit measurement}
\begin{figure*}[h!]
    \centering
    \begin{subfigure}[b]{0.60\linewidth}
        \centering
        \includegraphics[width=\textwidth]{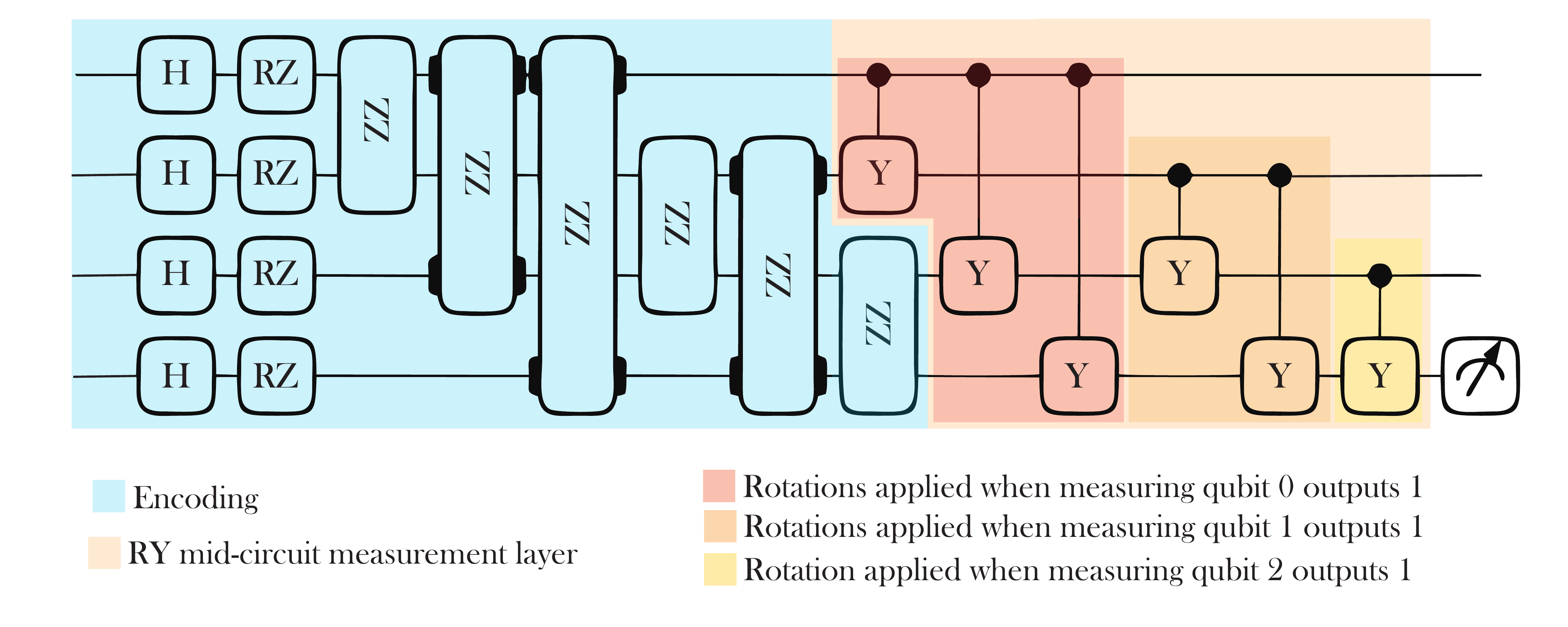}
        \caption{Mid-circuit measurement pooling example. This circuit consists of a higher-order encoding, followed by a $R_{Y}$ mid-circuit measurement layer.}
        \label{2a}
    \end{subfigure}
    \hfill
    \begin{subfigure}[b]{0.70\linewidth}
        \centering
        \includegraphics[width=\textwidth]{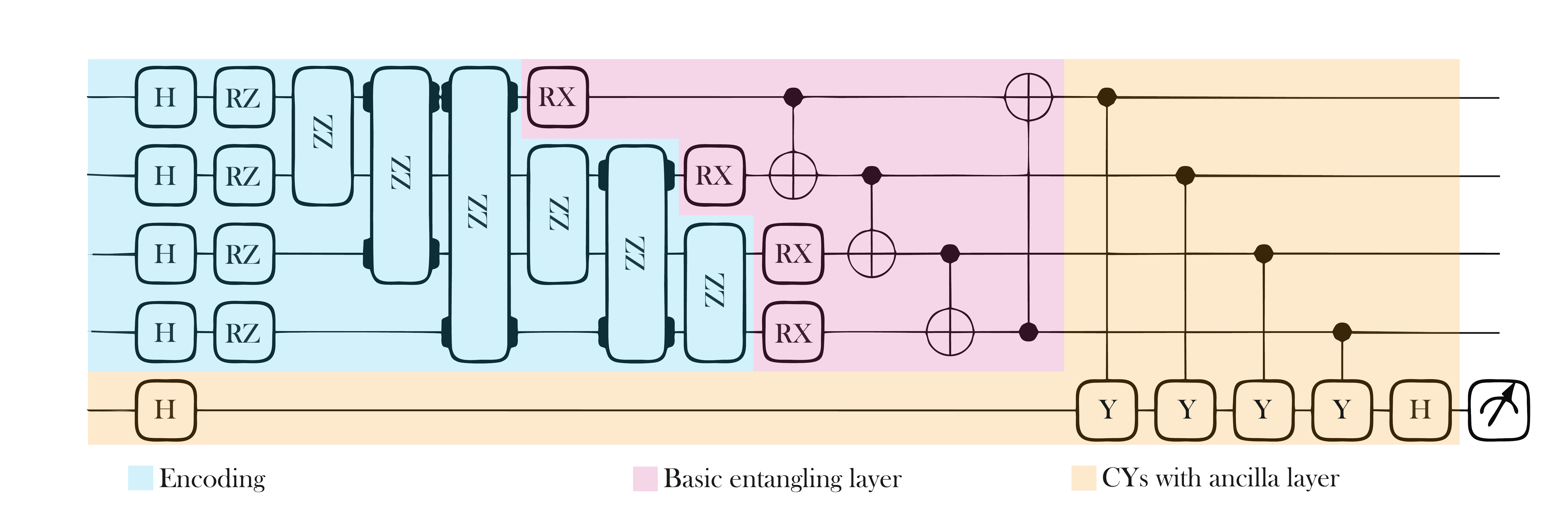}
        \caption{Ancilla qubit with controlled gates pooling example. This circuit consists of a higher-order encoding, followed by a basic entangling layer and CYs with ancilla qubit.}
        \label{2b}
    \end{subfigure}
    \hfill
    \begin{subfigure}[b]{0.60\linewidth}
        \centering
        \includegraphics[width=\textwidth]{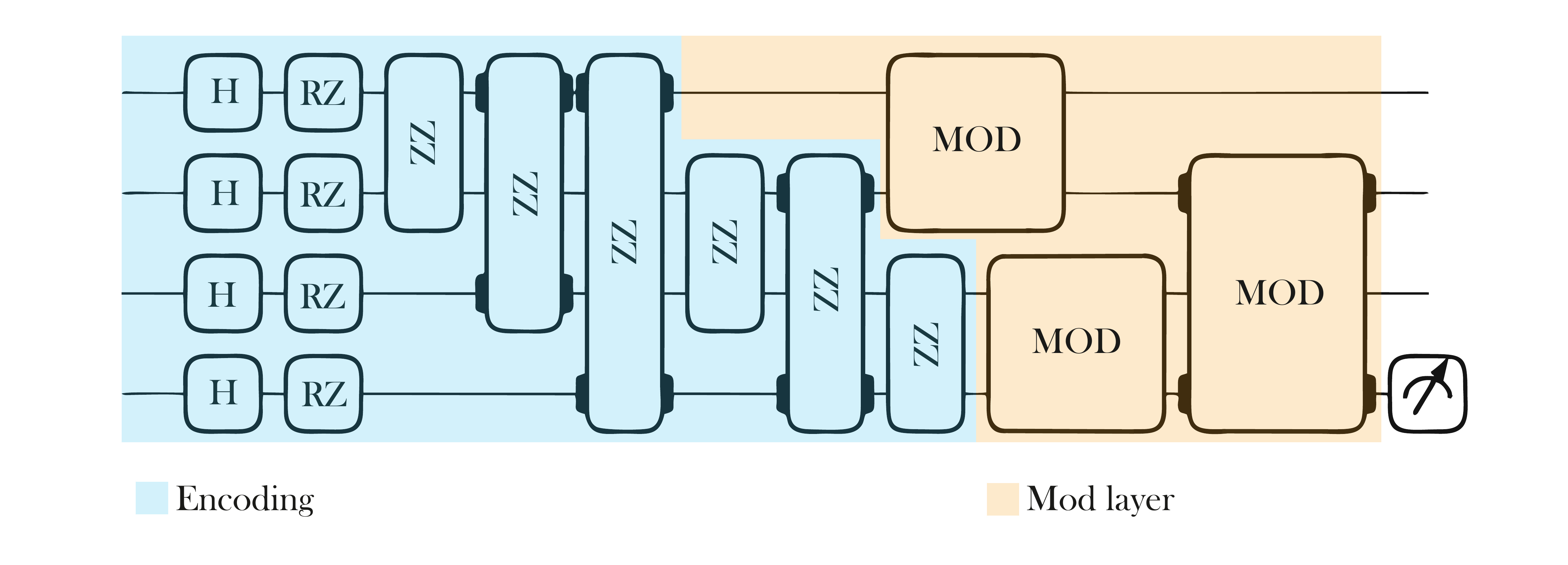}
        \caption{Modular quantum pooling example. This circuit consists of a higher-order encoding, followed by modular quantum pooling blocks. The alternatives for the blocks are given in (d), (e) and (f).}
        \label{2c}
    \end{subfigure}\\
    \begin{subfigure}{0.27\textwidth}
    \includegraphics[width=\textwidth]{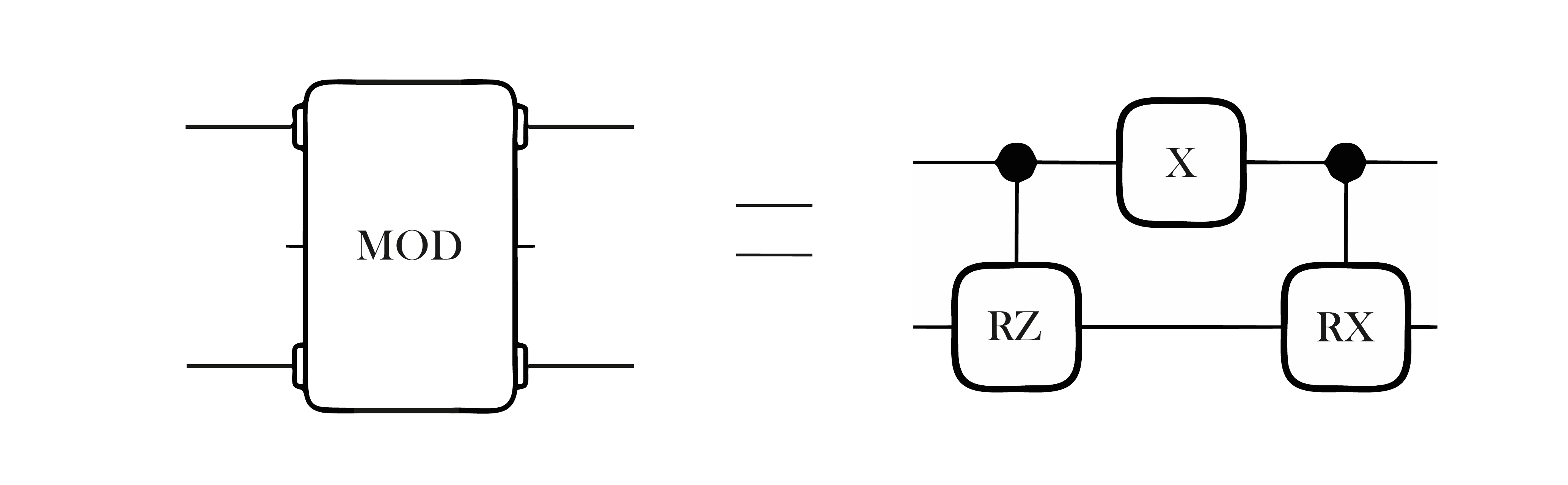}
    \caption{Mod-a}
    \label{2d}
    \end{subfigure}\hfill
    \begin{subfigure}{0.31\textwidth}
    \includegraphics[width=\textwidth]{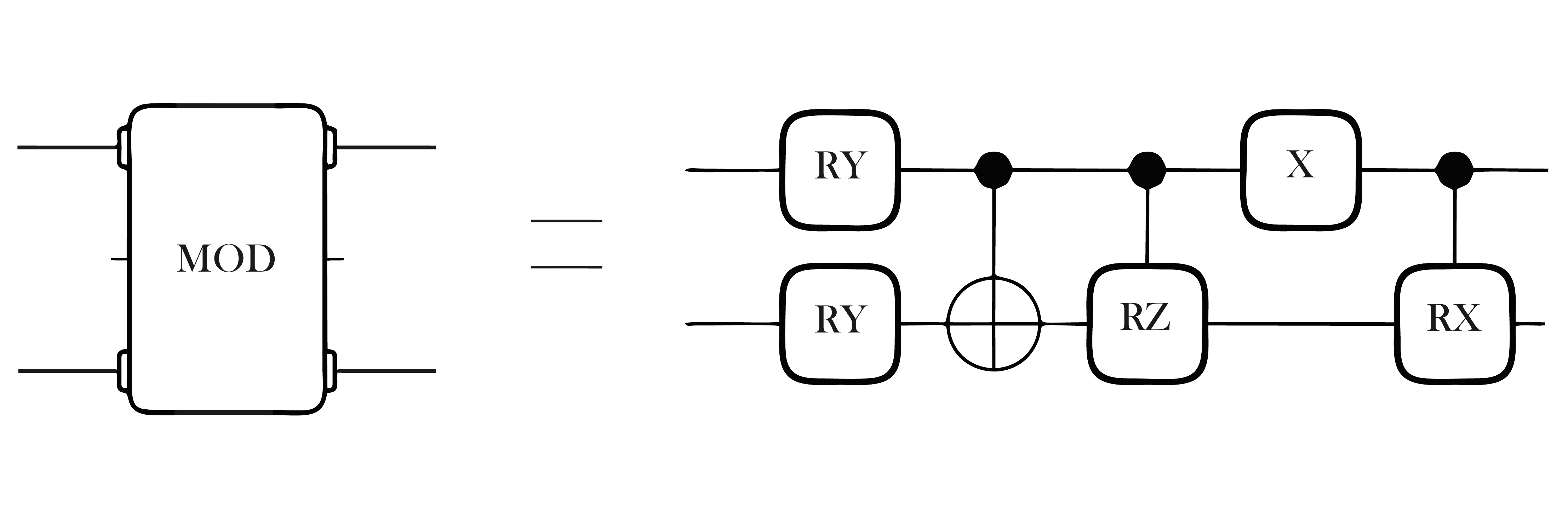}
    \caption{Mod-b}
    \label{2e}
    \end{subfigure}\hfill
    \begin{subfigure}{0.35\textwidth}
    \includegraphics[width=\textwidth]{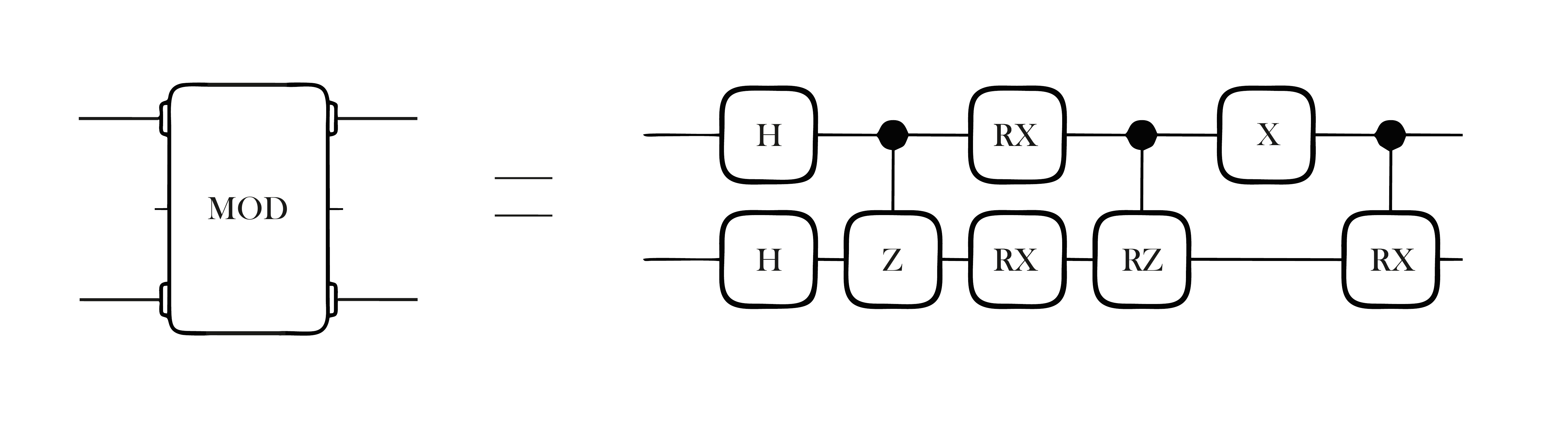}
    \caption{Mod-c}
    \label{2f}
    \end{subfigure}
    \hfill
    \begin{subfigure}[b]{0.70\linewidth}
        \centering
        \includegraphics[width=\textwidth]{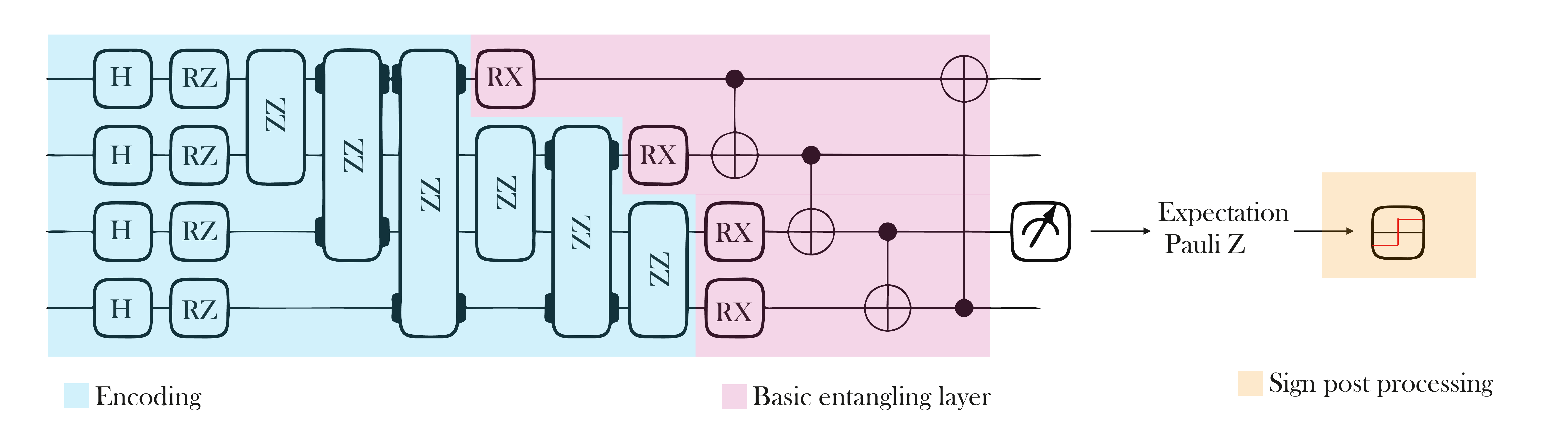}
        \caption{Qubit selection with classical postprocessing pooling example. This circuit consists of a higher-order encoding, followed by a basic entangling layer and a qubit selection pooling with $Sign(x)$ postprocessing}
        \label{2g}
    \end{subfigure}
\caption{Architecture sketches of the tested pooling architectures.}
\end{figure*}

Mid-circuit measurements allow for the introduction of additional parameters while reducing circuit depth, offering a favorable trade-off between parameter count and circuit complexity. In the present study, mid-circuit measurements are used to control multiple qubits simultaneously - equivalent to a multi-qubit controlled gate. This is particularly advantageous because implementing multi-qubit controlled gates in practice can be challenging due to current hardware limitations. As such, this technique offers a more efficient and practical alternative to multi-qubit controlled gates with several hardware devices being able to perform mid-circuit measurements \cite{govia2022randomized}. 

Mid-circuit measurements are used to implement alternative paths depending on the measured outcome. Given four qubits $q_0$, $q_1$, $q_2$, and $q_3$, qubit $q_0$ is measured and the rotation gates $R_{\theta_0}$, $R_{\theta_1}$, and $R_{\theta_2}$ are applied to qubits $q_1$, $q_2$, and $q_3$ respectively if the measurement outcome of qubit $q_0$ is 1. Similarly, qubit $q_1$ is subsequently measured and if the measurement yields 1, the rotations gates $R_{\theta_3}$ and $R_{\theta_4}$ are applied to qubits $q_2$ and $q_3$. Finally, a CNOT gate between $q_2$ and $q_3$ is applied, $q_2$ is measured and the rotation $R_{\theta_5}$ is applied to qubit $q_3$ if the measurement yields 1. The last qubit $q_3$ is measured and its value is fed to the classical fully-connected layer. The corresponding circuit is represented in figure \ref{2a}. Two different options using rotation gates with trainable angles $R_{X}$ and $R_{Y}$ are tested within this pooling method.

\begin{figure*}
\centering
 \begin{subfigure}{0.4\textwidth}
    \centering
    \includegraphics[width=\textwidth]{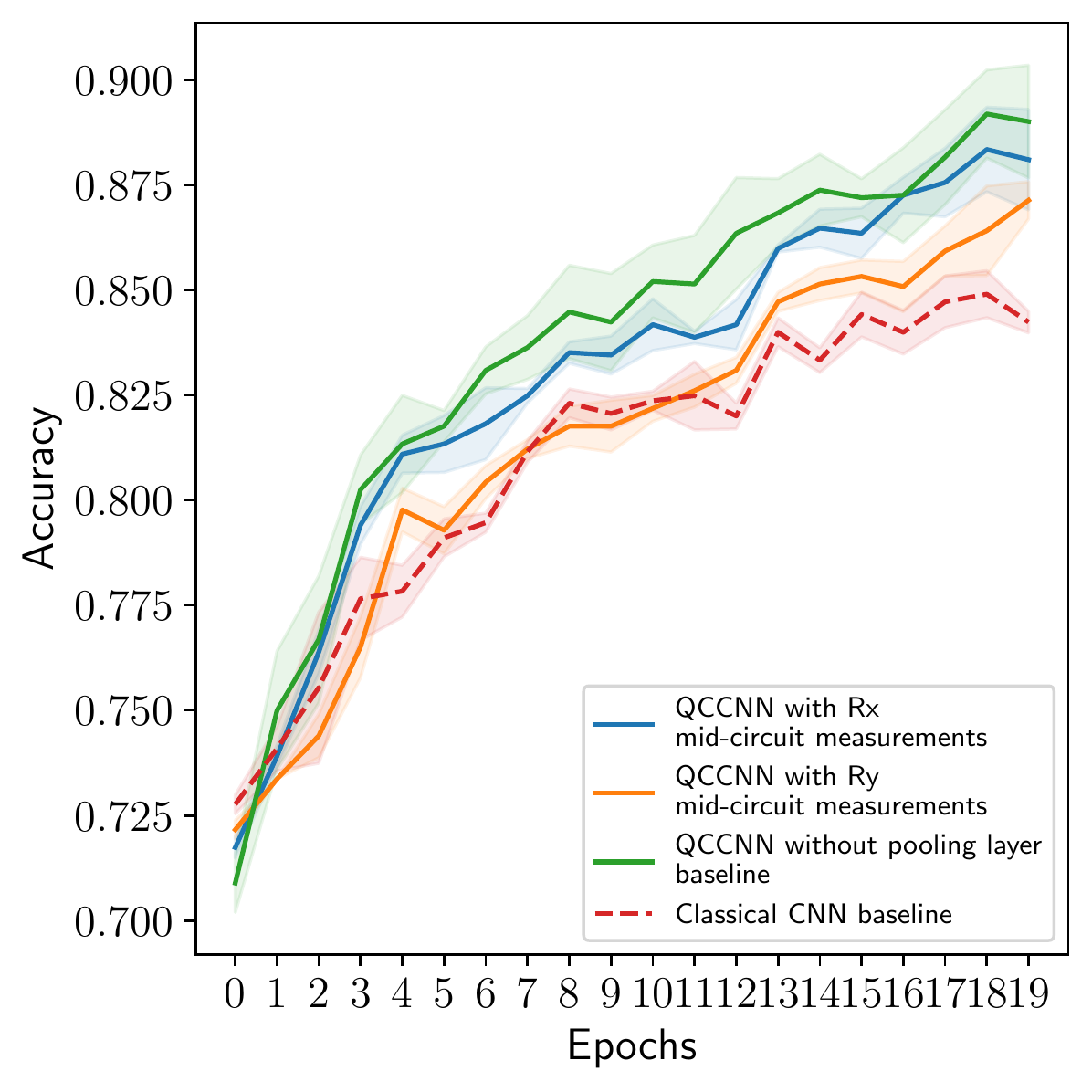}
    \caption{Training accuracy}
 \end{subfigure}
 \qquad
 \begin{subfigure}{0.4\textwidth}
    \centering
    \includegraphics[width=\textwidth]{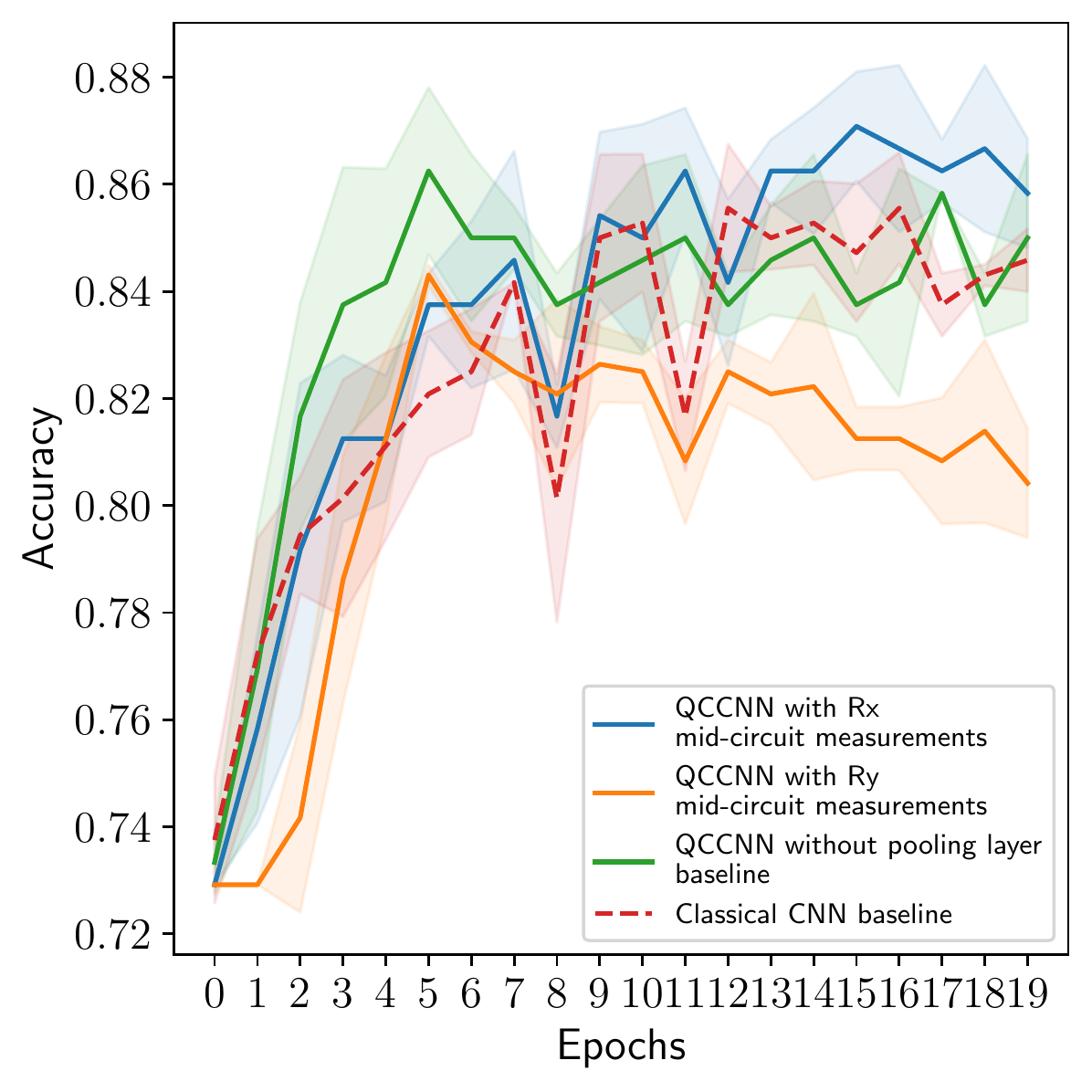}
    \caption{Validation accuracy}
\end{subfigure}
\qquad
\caption{Comparison of the performance of mid-circuit measurement pooling methods for hybrid QCCNNs in terms of training
and validation accuracy.}
\label{resultsmid}
\end{figure*}

The training and validation accuracy curves are presented in figure~\ref{resultsmid} and directly compared with the classical CNN baseline and the QCCNN baseline without pooling. We first remark that the training accuracy of the hybrid quantum-classical variants all surpass the classical model. When generalizing on the validation set, the architecture with $R_{X}$ mid-circuit measurements clearly performs better than the architecture with $R_{Y}$ mid-circuit measurements. It also surpasses the two baselines in maximum validation accuracy, although the overlapping variance bands prevent us to conclude on its superiority.

Additional experiments were run with a bigger learning rate of 0.01, that are not shown in the results curves for sake of comparison. With this learning rate, although overfitting occurs very fast, $R_{X}$ mid-circuit measurements is the only method presented within this paper where the training data was perfectly learnt with a training accuracy reaching 1.0 (validation accuracy of 88.47\%). These results show that pooling methods with mid-circuits measurements have a high potential in extracting relevant features for the training of QCCNNs.

\subsection{Ancilla qubit and controlled gates}
\begin{figure*}
\label{figancilla}
\end{figure*}

Models with no inductive bias are likely to have trainability and generalization issues, therefore group-invariant quantum machine learning models have been proposed and one of their fundamental component is the use of controlled gates and ancillary qubits \cite{larocca2022group}. We draw upon these models and use a standard structure when using an ancilla qubit, i.e., Hadamard-controlled gates-Hadamard, where in this work the controlled gates are either CY or CZ.

Our experiment involves initializing the system in a state consisting of four data qubits and one ancilla (control) qubit. Subsequently, the ancilla qubit is subjected to the Hadamard gate $H$. The input data is encoded on the first four qubits and a basic entangling layer is used. Following this, a series of four controlled gates is executed, wherein a controlled gate (either CY or CZ) is utilized with qubit $q_0$ as the control and the ancilla as the target, followed by a controlled gate with qubit $q_1$ as the control and the ancilla as the target, and so forth for qubits $q_2$ and $q_3$. Upon completion of these operations, the Hadamard gate $H$ is applied to the ancilla qubit. Finally, the Pauli-Z expectation value of the ancilla qubit is taken as the output of the circuit. The architecture of the ancilla qubit with CY controlled gates pooling is shown on figure \ref{2b}.

\begin{figure*}
\centering
 \begin{subfigure}{0.4\textwidth}
    \centering
    \includegraphics[width=\textwidth]{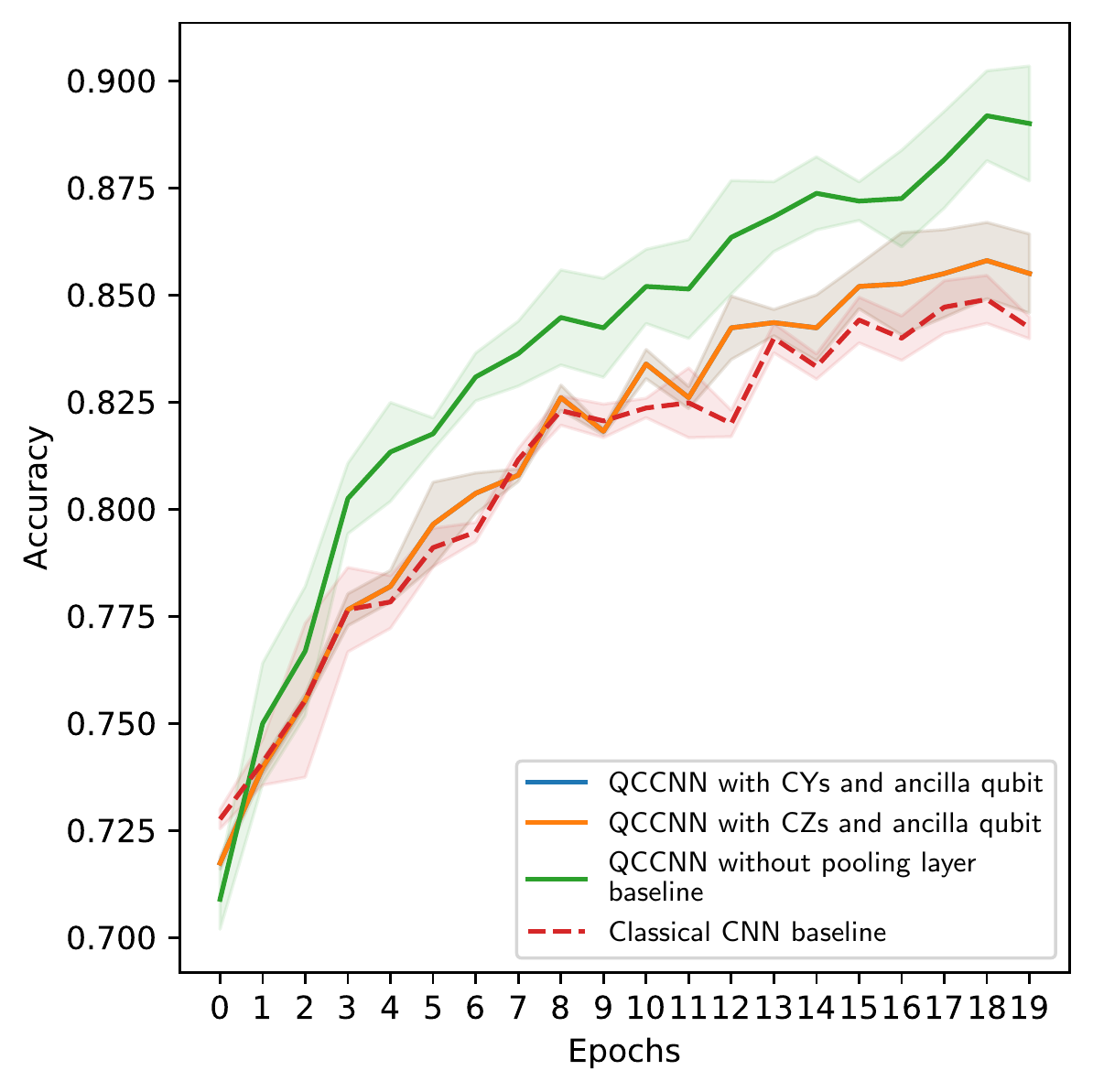}
    \caption{Training accuracy}
 \end{subfigure}
 \qquad
 \begin{subfigure}{0.4\textwidth}
    \centering
    \includegraphics[width=\textwidth]{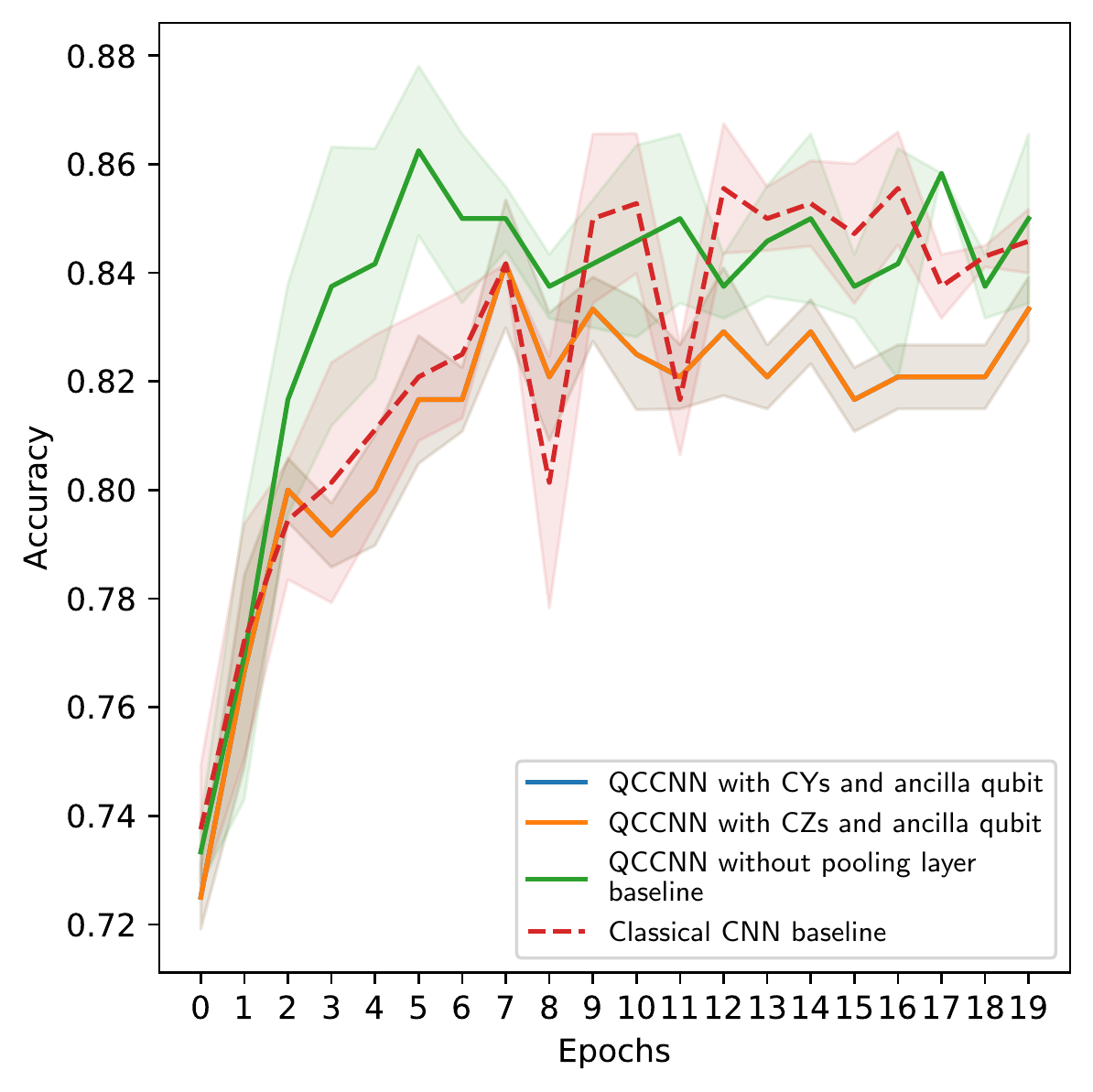}
    \caption{Validation accuracy}
\end{subfigure}
\qquad
\caption{Comparison of the performance of ancilla qubit and controlled gates pooling methods for hybrid QCCNNs in terms of training and validation accuracy.}
\label{resultancilla}
\end{figure*}

Our results in figure \ref{resultancilla} show that the use of CY or CZ controlled gates does not affect the overall performance of the QCCNN, as the curves for both variants strictly overlap. Overall, the ancilla qubit with controlled gates pooling method seems to have a satisfactory training performance, reaching a maximum accuracy slightly above the classical CNN. However, this method does not perform well in validation accuracy, with a maximum validation accuracy lower than both classical CNN and QCCNN without pooling baselines.

\subsection{Modular quantum pooling blocks}

\begin{figure*}
\label{figpool}
\end{figure*}

\begin{figure}[t]
\end{figure}

Taking inspiration from \cite{hur2022quantum}, we explore the alternative use of a combination of modular small quantum convolutional and pooling operations.
The circuit architecture of this quantum pooling layer is displayed in figures \ref{2c} to \ref{2f}. The design of this pooling method incorporates modular blocks acting on two neighboring qubits (on the first and second qubit and on the third and fourth qubit). Each of these blocks either consists of just a quantum pooling operation (Mod-a), or of a quantum convolutional circuit in addition to a quantum pooling operation. For the convolutional part, two different architectures are investigated, where the first variant is inspired by tree tensor network ideas (Mod-b), while the second circuit variant is known to have good entanglement properties \cite{sim2019expressibility} (Mod-c). As a result of the first two pooling blocks, a dimensionality reduction from four to two qubits occurs. The other qubits were traced out during the pooling operation. The remaining two qubits undergo another convolutional and pooling operation, using the same architecture as in the blocks used earlier in the circuit. Consequently, the whole quantum circuit performed a dimensionality reduction from four to one qubit. The remaining qubit is measured and then fed into the following classical parts in the network. Due to the modular design of this circuit architecture, the concept is easily extendable and adaptable to more complicated scenarios.

\begin{figure*}
\centering
 \begin{subfigure}{0.4\textwidth}
    \centering
    \includegraphics[width=\textwidth]{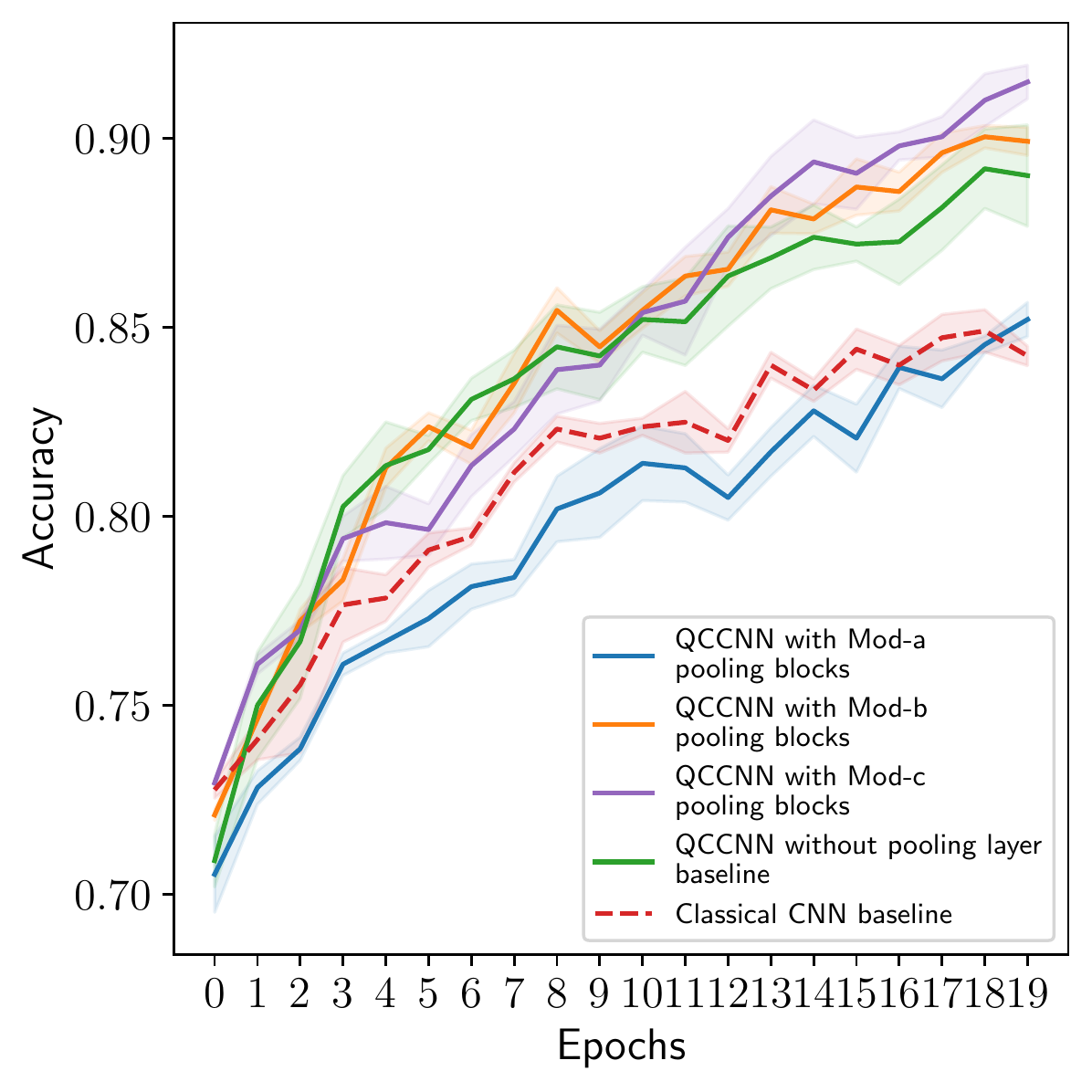}
    \caption{Training accuracy}
 \end{subfigure}
 \qquad
 \begin{subfigure}{0.4\textwidth}
    \centering
    \includegraphics[width=\textwidth]{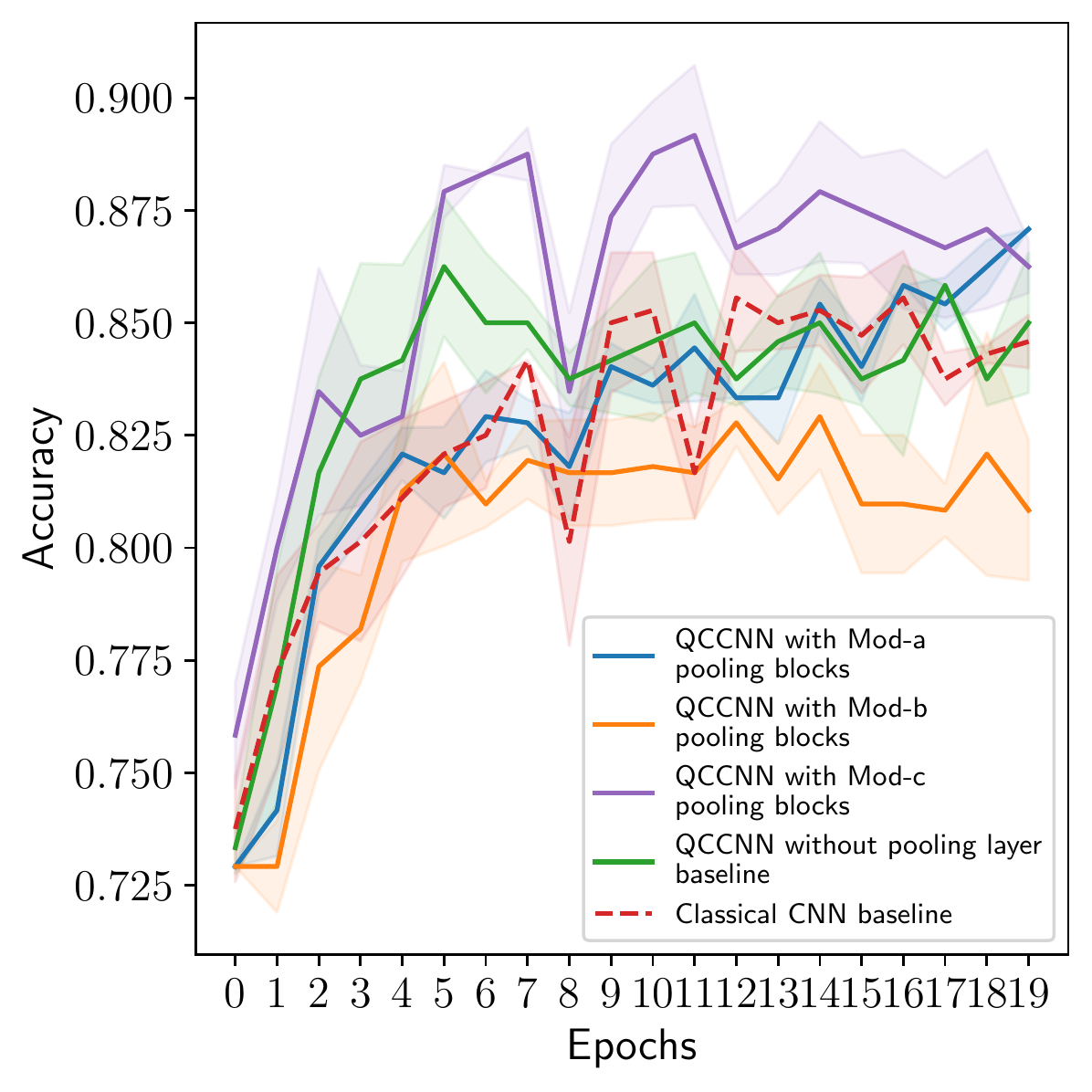}
    \caption{Validation accuracy}
\end{subfigure}
\qquad
\caption{Comparison of the performance of modular quantum pooling blocks for hybrid QCCNNs in terms of training and validation accuracy.}
\label{resultsmod}
\end{figure*}

We obtain in general very good performance with the modular pooling, with two methods out of three outperforming the classical baseline in validation accuracy, namely Mod-a and Mod-c, and all three methods surpassing the classical baseline in training accuracy, as seen on figure~\ref{resultsmod}. Particularly, the modular pooling method with Mod-c presents a substantial gain in performance compared to the classical baseline, with a fast and steep convergence early on in the training and clear and non overlapping error bands with the other models.

\subsection{Qubit selection with classical postprocessing}
\begin{figure*}
\end{figure*}

Inspired by the work of \cite{schuld}, after encoding the data and using a basic entangling layer, one out of the four qubits from the 4-qubit system is selected, onto which a classical postprocessing step is applied. This qubit could either remain fixed throughout the computation or be randomly selected for each iteration. We choose to measure qubit $q_2$ in our experiments. With this last pooling method, it can be investigated whether quantum circuits profit from non-linear activation functions in the same way as classical algorithms do. The output of all tested circuits being an expectation value between -1 and 1, the $Sign(x)$ and $Tanh(x)$ non-linear activation functions are tested, thereby ensuring that the output of the quantum circuit remains within the same range. The circuit is shown in figure \ref{2g}.
The $Sign(x)$ function is defined as
$$ Sign(x)  = \begin{cases} -1, & x < 0, \\ 0, & x = 0, \\ 1, & x > 0, \end{cases} $$

and $Tanh(x)$ is defined as
$$Tanh(x) = \frac{sinh(x)}{cosh(x)} = \frac{e^x - e^{-x}}{e^x + e^{-x}}.$$

\begin{figure*}
\centering
 \begin{subfigure}{0.4\textwidth}
    \centering
    \includegraphics[width=\textwidth]{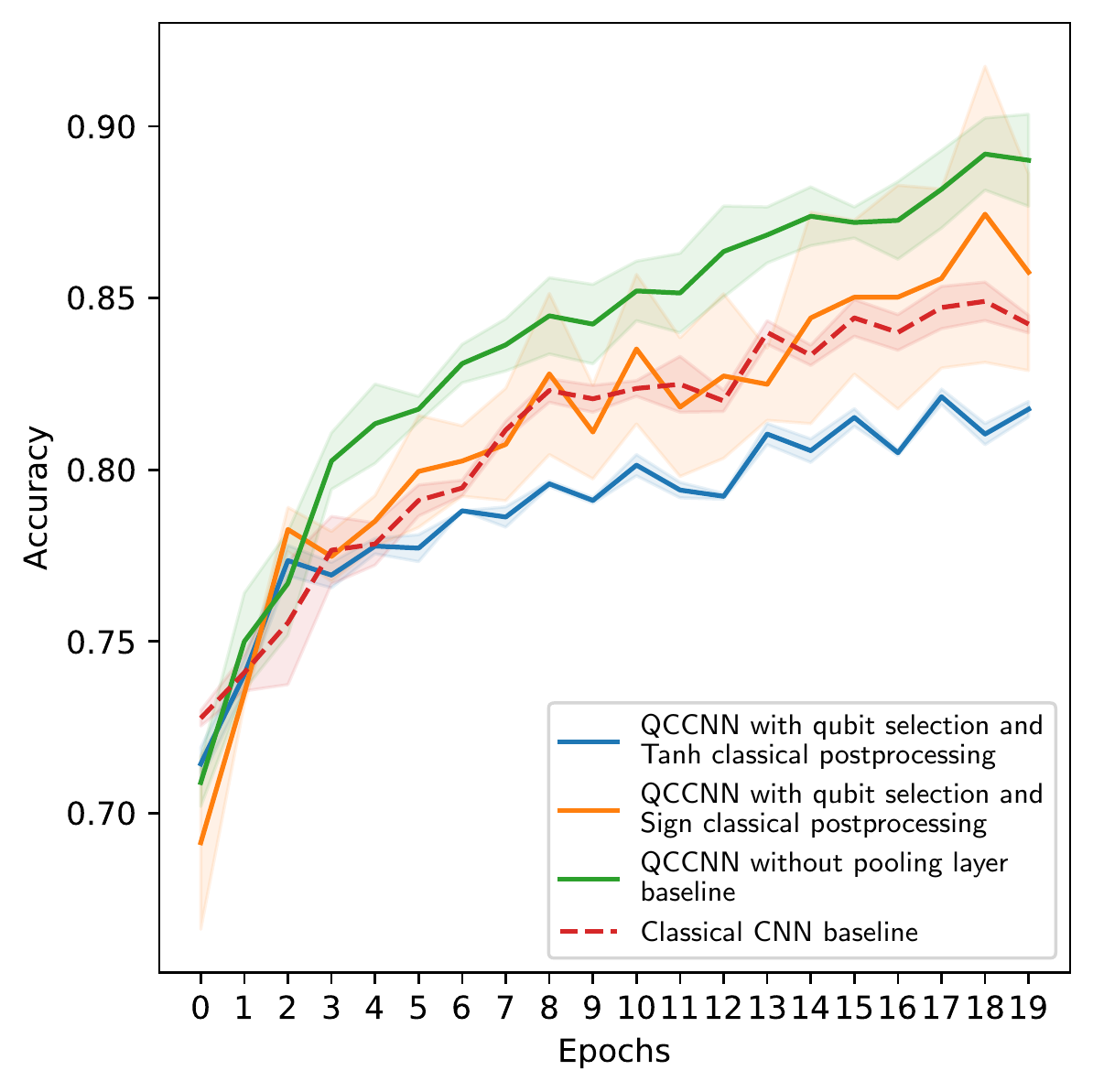}
    \caption{Training accuracy}
 \end{subfigure}
 \qquad
 \begin{subfigure}{0.4\textwidth}
    \centering
    \includegraphics[width=\textwidth]{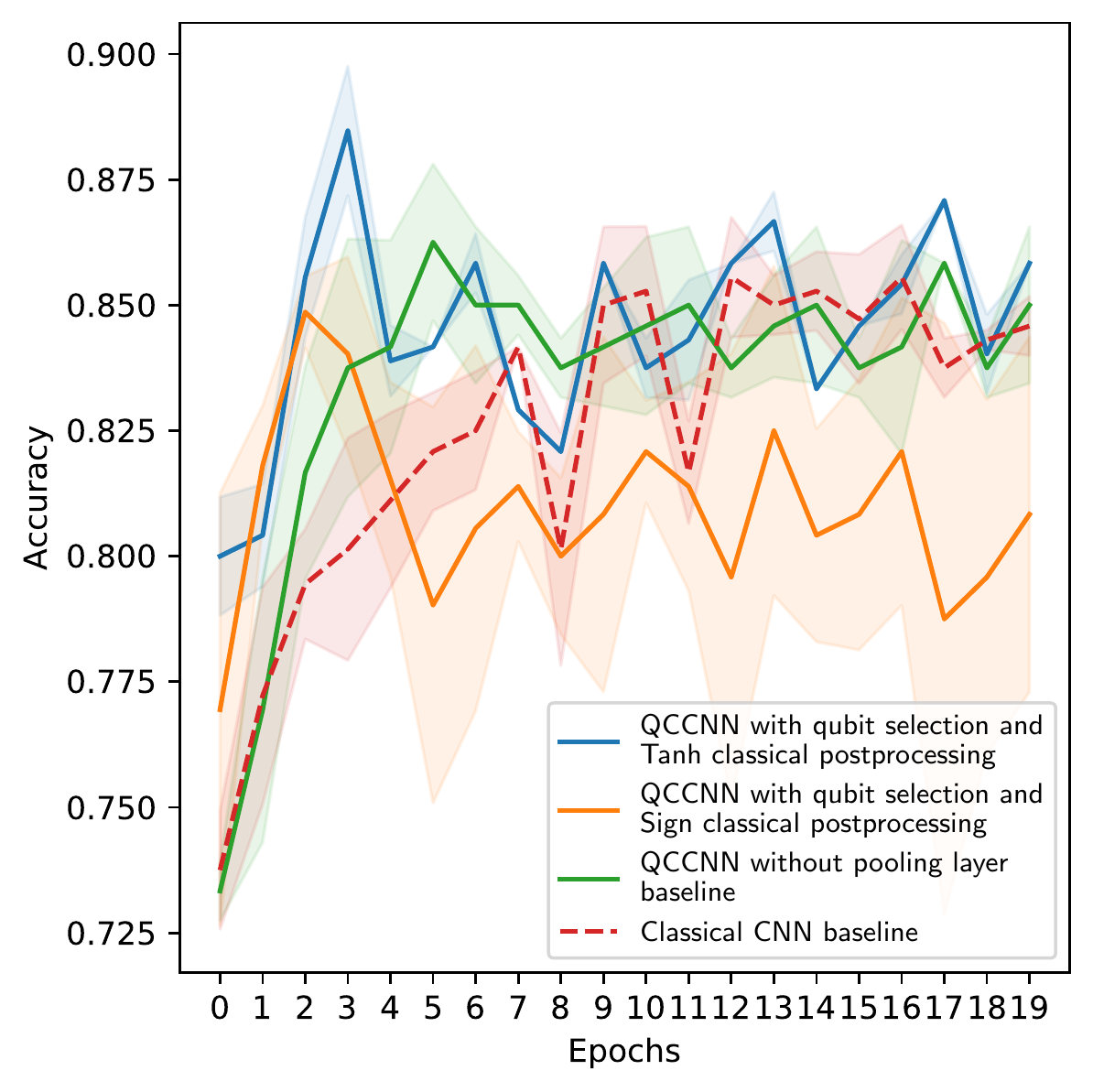}
    \caption{Validation accuracy}
\end{subfigure}
\qquad
\caption{Comparison of the performance of qubit selection with classical postprocessing pooling methods for hybrid QCCNNs in terms of training and validation accuracy.}
\label{resultpost}
\end{figure*}

The results of the quantum pooling with these postprocessing functions are shown in figure~\ref{resultpost}. We find that the architecture with the $Tanh(x)$ activation is able to generalize very well and quickly on the validation data, outperforming both the classical baseline and the QCCNN without pooling at the third epoch in average validation accuracy. However, one notes that the variance in validation accuracy does not allow us to unreservedly decipher between the performance of this variant and the QCCNN without pooling. The architecture with the $Sign(x)$ activation function is on the contrary not able to generalize properly on the validation data and suffers from remarkably wide variance bands. This probably stems from the information loss when the $Sign(x)$ function is applied.

\subsection{Review of the best pooling options}
\begin{figure*}
\centering
 \begin{subfigure}{0.4\textwidth}
    \centering
    \includegraphics[width=\textwidth]{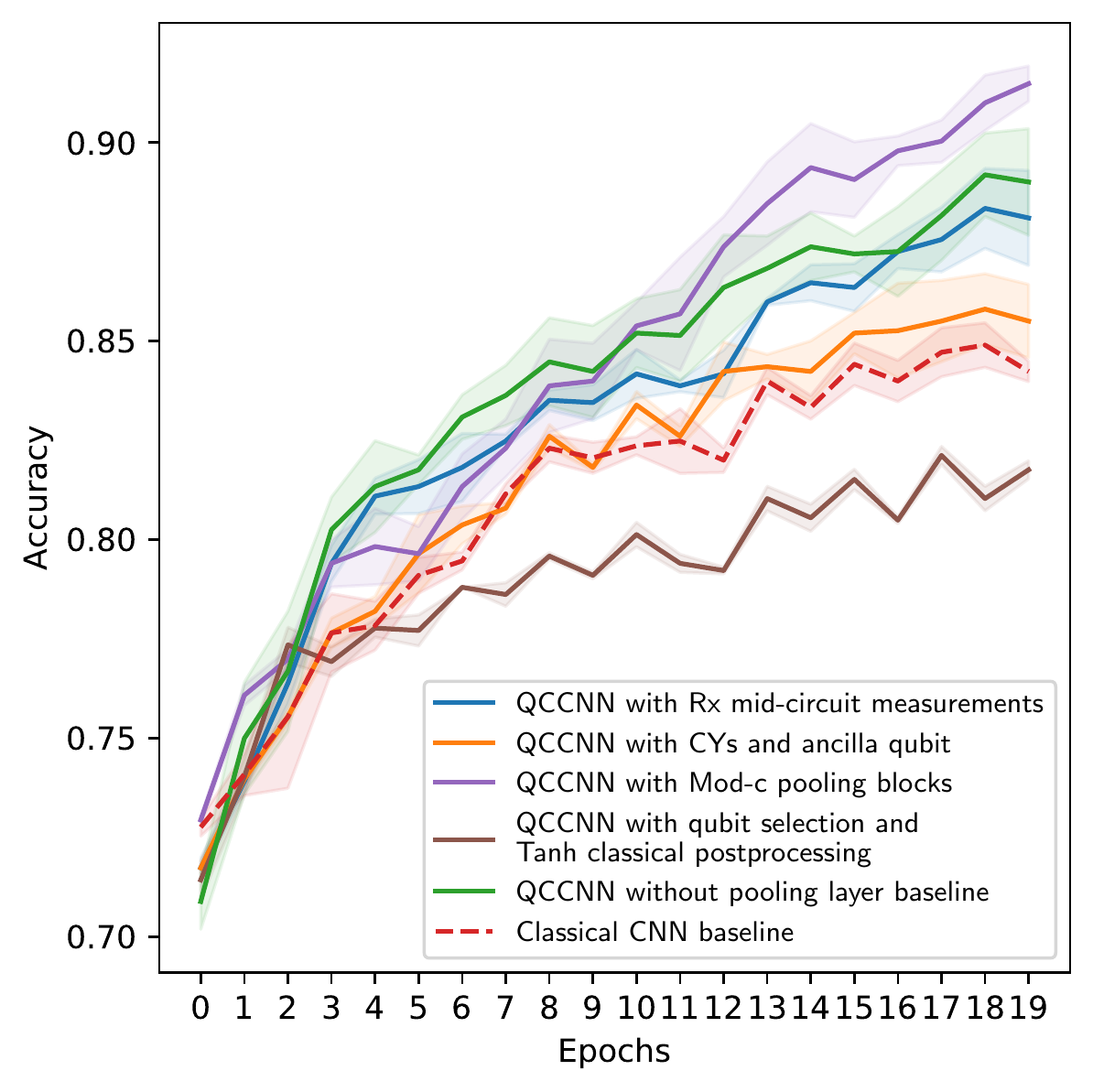}
    \caption{Training accuracy}
 \end{subfigure}
 \qquad
 \begin{subfigure}{0.4\textwidth}
    \centering
    \includegraphics[width=\textwidth]{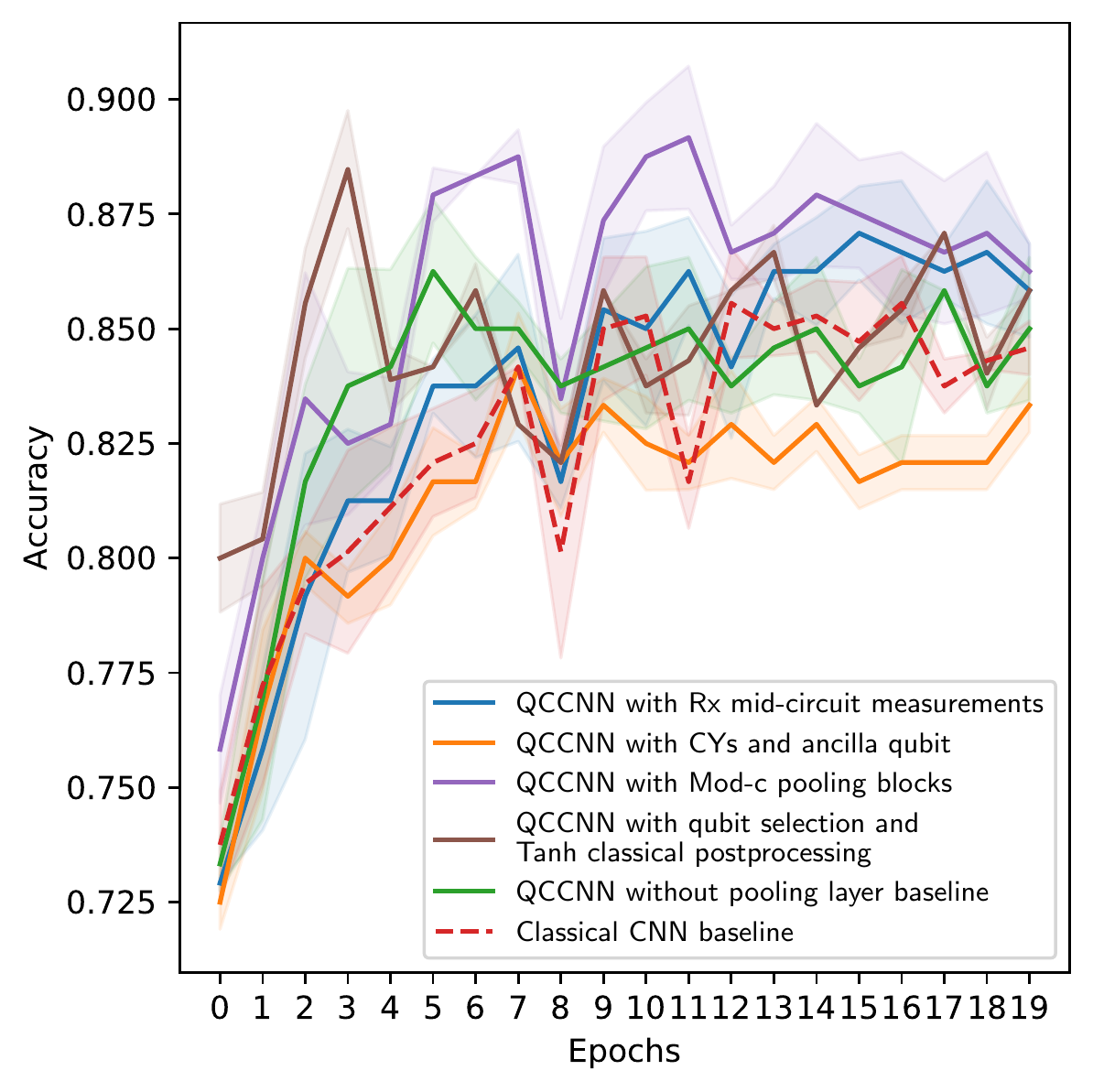}
    \caption{Validation accuracy}
\end{subfigure}
\qquad
\caption{Comparison of the best-performing pooling methods for hybrid QCCNNs in terms of training and validation accuracy.}
\label{resultsbest}
\end{figure*}

A summary plot is provided in figure \ref{resultsbest} showing the best performing configuration for each pooling option. It can be observed that most pooling architectures beat the classical CNN baseline in terms of the attained average maximal training and validation accuracy given a similar number of trainable parameters. One also notices a faster convergence for QCCNNs models in terms of validation accuracy as they achieve their highest validation accuracy sooner than the classical baseline, except for the mid-circuit measurements pooling method. These two observations are in line with what has been previously shown in theory \cite{Caro2021}. In some cases however, it is recommended to exercise some level of caution when interpreting these results due to the overlapping variance bands.

When comparing the different pooling methods together, the QCCNN with Mod-c modular pooling clearly performs best on the chosen dataset in both training and validation accuracy, also when taking into account the errors bands. This method is followed in maximum validation accuracy by the QCCNN with qubit selection with $Tanh(x)$ classical postprocessing, and the QCCNN with $R_{X}$ mid-circuit measurements, both options beating the QCCNN without pooling. Only the QCCNN with ancilla qubit pooling is performing less well than the baseline without pooling. In the next section, we attempt to give some insights on why some of these quantum circuits perform better than others.

\section{Effective dimension analysis}\label{section5}
In this final section, we aim to determine whether the performance of a specific VQC - and therefore a pooling method - could be foreseen using a quantum metric. Candidates for such a metric that were suggested in the literature are the effective dimension, the expressibility and the entanglement capability \cite{sim2019expressibility} \cite{abbas2021power}. However, the last two metrics assume a constant number of qubits in input and output of the VQC and were therefore not selected.

\subsection{The effective dimension}
The effective dimension (ED) is a complexity measure motivated by information geometry and designed to estimate the information capacity of the model. It can be computed for both classical and quantum neural networks by using the Fisher Information Matrix (FIM), a metric in statistics used to evaluate the impact of the variance of the models' parameters on its output. The FIM is given by
\begin{equation}
    F(\theta) = \mathbb{E} \left[ \frac{\partial}{\partial \theta} \log p(x,~y;~\theta)~ \frac{\partial}{\partial \theta} \log p(x,~y;~\theta)^T \right] \in \mathbb{R}^{d \times d},
\end{equation}
with $p(x,~y;~\theta)$ the conditional distribution describing the relationship between the input $x \in \mathbb{R}^{s_{in}}$, the output $y \in \mathbb{R}^{s_{out}}$, and the parameters $\theta \in \Theta$, where $\Theta \subset \mathbb{R}^d$ is the Riemannian space of the model parameters.

If one uses finite sampling, the FIM can be approximated empirically to
\begin{equation}
    \Tilde{F}_k(\theta) = \frac{1}{k} \sum_{j=1}^k \frac{\partial}{\partial \theta} \log p(x_j,~y_j;~\theta)~ \frac{\partial}{\partial \theta} \log p(x_j,~y_j;~\theta)^T,
\end{equation}
where $k$ is the number of independent and identically distributed (i.i.d) samples $(x_j, y_j)$ drawn from $p(x,~y;~\theta)$. This formulation based on \cite{berezniuk2020scale} was extended by \cite{abbas2021power} with the addition of the constant $\gamma \in (0,1]$ and a $\log n$ term to assure the ED is bounded. The ED is therefore defined as
\begin{equation}
    d_{\gamma, n}(\mathcal{M}_\Theta) = \frac{\log(\frac{1}{V_\Theta} \int_\Theta \sqrt{ \det (\text{id}_d + \frac{\gamma n}{2 \pi \log n} \hat{F}(\theta))}~d\theta)}{\log(\frac{\gamma n}{2 \pi \log n})}, 
\end{equation}
where $n > 1 \in \mathbb{N}$ is the number of data samples,  $V_\Theta \doteq \int_\Theta d\theta$ is the volume of the parameter space, and $\hat{F}(\theta)$ is the normalised FIM
\begin{equation}
    \hat{F}_{ij} (\theta)= d \frac{V_\Theta}{\int_\Theta tr(F(\theta))~d\theta} F_{ij}(\theta).
\end{equation}
The resulting ED is normalized by the number of training parameters for better comparison between the different VQCs.

\subsection{Metric and model performance}
Table \ref{table:results_all} shows the calculated ED of the VQCs for all tested quantum mechanisms as well as for the established QCCNN baseline without pooling. Note that we choose to focus on the VQC architecture in the calculation of the ED, thus the classical postprocessing function is not included for the qubit selection with classical postprocessing method.

The ED being a measure of the information capacity of the model, we expect the training accuracy to be positively correlated with this metric. However, this is not a conclusion that can be drawn within this work. Mid-circuit measurements have a relatively high normalized ED and indeed perform well in training accuracy. On the other hand, Mod-a modular pooling also has a good normalized ED but the performance in training was worse. Among the lowest normalized ED values is Mod-c, which is in fact the best performing variant. If not taking into account Mod-b and Mod-c pooling options, we could in fact observe a positive correlation between ED and performance of the VQC. We hypothesize that such a relation might still exist, and that it should be further explored in future work. It can be for example studied whether the effect of different optimizers could reveal the potential link between ED and VQC performance.

\begin{table*}[!h]
    \centering
    \caption{The ED of all trainable architectures reflected against the maximum training and validation accuracies obtained during training. The higher the ED, the better the VQC with respect to that metric.}
    \begin{tabular}[width=\textwidth]{| l | c | c | c | c | c | c | c | c |}
       \hline
        VQC & Normalized ED & Max training accuracy & Max validation accuracy \\
        \hline
        $R_{X}$ mid-circuit measurements & 0.909 $\pm$ 0.016 & 92.81 $\pm$ 0.37 & 87.08 $\pm$ 1.02 \\
        $R_{Y}$ mid-circuit measurements & 0.906 $\pm$ 0.008 & 87.14 $\pm$ 0.44 & 84.31 $\pm$ 0.20 \\
        \hline
        Ancilla qubit with CY controlled gates & 0.772 $\pm$ 0.041 & 85.81 $\pm$ 0.89 & 84.17 $\pm$ 1.18 \\
        Ancilla qubit with CZ controlled gates & 0.801 $\pm$ 0.042 & 85.81 $\pm$ 0.89 & 84.17 $\pm$ 1.18 \\
        \hline
        Modular pooling Mod-a & 0.726 $\pm$ 0.006 & 85.21 $\pm$ 0.45 & 87.08 $\pm$ 0.00 \\
        Modular pooling Mod-b & 0.149 $\pm$ 0.003 & 90.04 $\pm$ 0.30 & 82.92 $\pm$ 1.18 \\
        Modular pooling Mod-c & 0.249 $\pm$ 0.016 & 91.49 $\pm$ 0.44 & 89.17 $\pm$ 1.56 \\
        \hline
        Qubit selection for classical postprocessing & 0.666 $\pm$ 0.048 & 82.13 $\pm$ 0.23 & 88.47 $\pm$ 1.29 \\
        \hline
        Basic entangling layer without pooling & 0.933 $\pm$ 0.019 & 89.19 $\pm$ 1.05 & 86.25 $\pm$ 1.56 \\
        \hline
    \end{tabular}
    \label{table:results_all}
\end{table*}

\section{Conclusion}\label{ccl}

This work presented four different architectures - mid-circuit measurements, ancilla qubits with controlled gates, modular quantum pooling blocks and qubit selection with classical postprocessing - to realize quantum pooling in hybrid QCCNNs. For this purpose, the classical convolutional layer within a classical CNN was replaced by a quantum convolution and pooling layer. The use of QCCNNs is particularly interesting for situations with little training data, which is typically the case in classification tasks on medical images. Therefore, we studied the different proposed architectures on a small dataset featuring ultrasound images of the breast to classify lesions as benign or malign. We find that quantum pooling variants perform at least similar or even in some cases clearly overcome the equivalent classical CNN and hybrid QCCNN without pooling, while using a similar number of trainable parameters. On the selected dataset, the modular pooling technique beat all variants on both training and validation sets. The mid-circuit measurements pooling seems to be particularly promising in terms of training ability, while the qubit selection with classical postprocessing exhibits very good generalization ability. These results, obtained in simulation, are very promising for further studies of QCCNNs, as those might be suited for presently available NISQ hardware already, given the small number of qubits required and the shallow depth of the quantum circuits involved. The presented pooling techniques are naturally also relevant in the design choice for fully quantum architectures. However, to fully exploit the potential benefits, additional work to produce insights about which circuit architectures will result in benefits for a given use case should be pursued.

\section{Acknowlegements}

The project/research is supported by the Bavarian Ministry of Economic Affairs, Regional Development and Energy with funds from the Hightech Agenda Bayern.

\bibliographystyle{unsrt}
\bibliography{bibliography}

\end{document}